# Multitaper-Based Post-Processing of Compact Antenna Responses Obtained in Non-Anechoic Conditions

Mariusz Dzwonkowski[1,2][0000-0003-3580-7448], Adrian Bekasiewicz[1][0000-0003-0244-541X], and Slawomir Koziel[1,3][0000-0002-9063-2647]

[1] Faculty of Electronics, Telecommunications and Informatics, Gdansk University of Technology, Narutowicza 11/12, 80-233 Gdansk, Poland
[2] Department of Radiology Informatics and Statistics, Faculty of Health Sciences, Medical University of Gdansk, Tuwima 15, 80-210 Gdansk, Poland
[3] Department of Engineering, Reykjavik University, Menntavegur 1, 102 Reykjavík, Iceland
bekasiewicz@ru.is

**Abstract.** The process of developing antenna structures typically involves prototype measurements. While accurate validation of far-field performance can be performed in dedicated facilities like anechoic chambers, high cost of construction and maintenance might not justify their use for teaching, or low-budget research scenarios. Non-anechoic experiments provide a cost-effective alternative, however the performance metrics obtained in such conditions require appropriate correction. In this paper, we consider a multitaper approach for post-processing antenna far-field characteristics measured in challenging, non-anechoic environments. The discussed algorithm enhances one-shot measurements to enable extraction of line-of-sight responses while attenuating interferences from multi-path propagation and the noise from external sources of electromagnetic radiation. The performance of the considered method has been demonstrated in uncontrolled conditions using a compact spline-based monopole. Furthermore, the approach has been favorably validated against the state-of-the-art techniques from the literature.

**Keywords:** Antenna calibration; data post-processing; non-anechoic measurements; radiation pattern, multitaper.

## 1  Introduction

Antennas are traditionally validated in costly, specialized laboratories where strict control over the propagation environment is maintained in order to mitigate its effects on the measured performance of the prototype [1]-[4]. However, for budget-sensitive applications like training of students, underfunded research, or rapid iterative development conducted by start-up companies the use or lease of professional facilities might be economically impractical. Instead, the tests can be performed in in-door or outdoor conditions not intended for precise far-field experiments such as offices, hallways, courtyards, or parks [5]. Despite potential cost savings, the results obtained in such uncontrolled propagation conditions are of little to no use for interpretation of the antenna under test (AUT) real-world performance [5]-[7].

The challenges related to insufficient fidelity of non-anechoic measurements can be mitigated using appropriate post-processing methods. The latter ones belong to two main categories that include: (i) decomposition of transmitted signals and (ii) characterization of propagation environments [5]-[18]. Signal decomposition techniques boil down to extraction of the Line-of-Sight (LoS) transmission between the reference antenna (RA) and AUT based on time, or frequency domain analyses that employ suitable windowing kernels, or a truncated composition of basis functions [5]-[14]. The second group of methods focuses on extracting the effects of propagation environment on the RA-AUT system performance (e.g., based on comparative measurements conducted within the professional test sites) [15]-[17]. Their practical implementations include analyzing the AUT in multiple spatially separated locations [17], determining the noise floor [15], or estimating equivalent currents on a hull enclosing the radiator [18].

Unfortunately, the existing approaches face challenges that limit their applicability to measurements in truly uncontrolled environments. Issues include not only cognitive and problem-specific setup of correction parameters, but also validation using electrically large, high-gain (and hence featuring increased signal-to-noise ratio) antennas [16], [17]. Finally, the state-of-the-art methods are predominantly demonstrated in favorable propagation conditions compared to standard office rooms, such as anechoic chambers (ACs), semi-ACs, or even based on electromagnetic (EM) simulation environments [10]-[14]. From this perspective, the problem concerning reliable measurements of small antennas in non-anechoic conditions remains unsolved.

In this work, a framework for automatic correction of measurements performed in uncontrolled environments is considered. The method involves modification of the noise-distorted responses obtained in non-anechoic conditions using a series of discrete prolate spheroidal sequences (DPSS) so as to enhance the relevant fraction of the RA-AUT impulse response while suppressing the noise and interferences. The post-processing results are obtained as a convex combination of responses refined using locally optimal DPSS functions. The performance of the algorithm is demonstrated based on four experiments conducted in a non-anechoic test site (a standard office room) and oriented towards evaluating far-field characteristics of a geometrically small spline-based monopole. Potential applications of the considered antenna include in-door localization, or radiology with emphasis on microwave imaging devices dedicated to breast cancer detection [19], [20]. Furthermore, the method is compared against the state-of-the-art techniques from the literature.

## 2      Methodology

In this section, a multitaper-based post-processing is discussed, starting with a formulation of the problem related to refining non-anechoic measurements. Subsequently, the correction procedure utilizing DPSS taper functions is outlined, followed by an explanation of the parametric tuning method applied to the considered framework.

### 2.1    Problem Formulation

Assume the set of uncorrected transmission responses (distorted by EM noise and multi-path interferences) of the RA-AUT system is denoted as $\boldsymbol{R}_u(\boldsymbol{\omega}, \boldsymbol{\theta})$. The parameters $\boldsymbol{\omega} = [\omega_1 \dots \omega_k \dots \omega_K]^T$ and $\boldsymbol{\theta} = [\theta_1 \dots \theta_a \dots \theta_A]^T$ refer to frequency sweep around $f_0 = (\omega_K - \omega_1)/2$ and angular rotation of AUT w.r.t. RA, respectively. The assumed bandwidth around $f_0$ is $B = \omega_K - \omega_1$. The multitaper post-processing method entails spectral analysis of $\boldsymbol{R}_u(\boldsymbol{\omega}, \boldsymbol{\theta})$ to obtain $\boldsymbol{R}_c^{*} = \boldsymbol{R}_c^{*}(f_0, \boldsymbol{\theta})$, i.e. the refined response of AUT at the frequency of interest $f_0$. The vector $\boldsymbol{R}_c^{*}$ can be interpreted as an estimate of the measurements carried out in an anechoic environment [1].

### 2.2    Multitaper Post-Processing

The post-processing relies on spectral analysis of the signal, incorporating multiple modifications of the uncorrected signals using a composition of mutually orthogonal taper functions, i.e. DPSSs. The correction process is as follows [7].

Assuming a specific angle $\theta_a$ for RA-AUT system, let the time-domain response be denoted as $\boldsymbol{T}_u = \boldsymbol{T}_u(\boldsymbol{t}, \theta_a) = F^{-1}(\boldsymbol{R}_u, N)$ and extracted via an inverse Fourier transform $F^{-1}(\cdot)$ with $N = 2^{\lceil \log_2 K \rceil + 3}$ points from $\boldsymbol{R}_u = \boldsymbol{R}_u(\boldsymbol{\omega}, \theta_a)$, where $\lceil \cdot \rceil$ denotes rounding up to the nearest integer. Additionally, let $\boldsymbol{t} = [t_1, \dots, t_N]^T = \partial t \cdot \boldsymbol{M}$ be the sweep in time for $\partial t = (\omega_K - \omega_1) \cdot (K-1)/(N-1)$ and $\boldsymbol{M} = [-N/2, \dots, N/2-2, N/2-1]^T$. Efficient noise isolation from the relevant fraction of the RA-AUT transmission can be achieved through multiple tapering of the segmented $\boldsymbol{T}_u$ with partial overlap of its subsequent intervals (segments) [21]. To segment $\boldsymbol{T}_u$ with an overlap, let $\boldsymbol{T}_s = \boldsymbol{T}_u(\boldsymbol{t}_s, \theta_a)$ be the $i$-th segment of $\boldsymbol{T}_u$, such that $\boldsymbol{t}_s = \partial t \cdot \boldsymbol{M}_s$, where $\boldsymbol{M}_s = \boldsymbol{M}_s(i) = [-N/2+(i-1)\cdot s, \dots, -N/2+(i-1)\cdot s+n-1]^T$. The parameter $i = 1, 2, \dots, \lfloor (N-n+s)/s \rfloor$ (where $\lfloor \cdot \rfloor$ indicates rounding down to the nearest integer) is determined by segment length $n = 1, 2, \dots, N-1$ and step length $s = 1, 2, \dots, n$, both defined in points. To achieve a reasonable $\boldsymbol{T}_u$ segmentation with a desirable overlap of consecutive $\boldsymbol{T}_u$ fractions, assume $1 \le s \le n \le N/2$.

The multiple tapering of segments from $\boldsymbol{T}_u$ requires the prior formulation of the specific DPSS functions. These can be obtained by finding all $n$-point finite energy sequences (eigenvectors) that maximize the spectral concentration ratio (corresponding eigenvalues) for the selected $2W$ bandwidth, where $W < 0.5 \partial t^{-1}$ [21], [22]. The procedure of finding all DPSS components $\boldsymbol{T}_\kappa(\boldsymbol{t}_s, w)$ for the $i$-th segment of $\boldsymbol{T}_u$ can be depicted as:

$$\sum_{\rho=0}^{n-1} \frac{\sin(2\pi(\gamma - \rho) \cdot W)}{\pi(\gamma - \rho)} \cdot \boldsymbol{T}_\kappa(\boldsymbol{t}_s(\rho), w) = \lambda_w \cdot \boldsymbol{T}_\kappa(\boldsymbol{t}_s(\gamma), w) \qquad (1)$$

where $\gamma = 0, 1, \dots, n-1$. Parameter $w = 0, 1, \dots, n-1$ is the order of the identified $\boldsymbol{T}_\kappa(\boldsymbol{t}_s, w)$ sequence for the spectral concentration ratio $\lambda_w$. Note that sequences of lower order have higher concentration ratios. The optimal number of DPSS tapers (with high enough $\lambda_w$) can be denoted as $w_{opt} = \lfloor 2t_{HB} \rfloor - 1$, where $t_{HB} = n \cdot \partial t \cdot W < n/2$ is a time-half-bandwidth product [22].

The identification of the DPSS functions enables modification of the segments from $\boldsymbol{T}_u$ using the obtained set of tapers, where each segment is modified using $w_{opt}$ functions. Subsequently, the set of $n$-point Fourier transforms of the individual tapered events are computed, and their convex combination, using $\lambda_w$ weights, is calculated to extract the frequency-domain response for each segment. Assuming the $i$-th segment $\boldsymbol{T}_s$, this task can be formulated as follows:

$$\boldsymbol{R}_{cs}(\boldsymbol{\Omega}_s,\ \theta_a) = \sum_{w=0}^{w_m-1} F(\boldsymbol{T}_s \circ \boldsymbol{T}_\kappa(\boldsymbol{t}_s,\ w),\ n) \cdot \frac{\lambda_w}{\sum_{w'=0}^{w_m-1} \lambda_{w'}} \tag{2}$$

where $\circ$ is the component-wise multiplication. The vector $\boldsymbol{R}_{cs}(\boldsymbol{\Omega}_s, \theta_a)$ represents the frequency-domain response for the $i$-th segment $\boldsymbol{T}_s$, such that $\boldsymbol{\Omega}_s = \partial\omega \cdot \boldsymbol{M}_s$ and $\partial\omega = (t_N - t_1)^{-1}$. Upon calculation, the frequency responses for all subsequent segments of $\boldsymbol{T}_u$ can be concatenated (with an applied arithmetic mean for overlapping points) to yield the overall frequency-domain response $\boldsymbol{R}_c(\boldsymbol{\Omega}, \theta_a)$. Here, $\boldsymbol{\Omega} = \partial\omega \cdot \boldsymbol{M}_\omega$, and $\boldsymbol{M}_\omega = [-N/2, \ldots, -N/2 + (\lfloor(N-n+s)/s\rfloor-1)\cdot s + n - 1]^T$. The corrected response of the RA-AUT system in non-anechoic conditions, i.e. $R_c(f_0, \theta_a)$, is obtained at $f_0 \in \boldsymbol{\Omega}$ and $\theta_a$ angle. This procedure is repeated for all $\theta_a$ angles to determine $\boldsymbol{R}_c(f_0, \boldsymbol{\theta})$.

### 2.3    Optimization of Multitaper Functions

The discussed multitaper post-processing is optimized using two parameters, i.e., segment length $n$ and step length $s$, both crucial for ensuring high correction performance when using DPSS functions. The optimal values for $n$ and $s$ (for the assumed $t_{HB} = 4$) are determined based on evaluations of a scalar objective function [7]:

$$U(\boldsymbol{x}) = \sum \left(\boldsymbol{R}_c(\boldsymbol{x}) - \alpha\boldsymbol{R}_f(f_0, \boldsymbol{\theta})\right) \tag{3}$$

where $\boldsymbol{R}_c(\boldsymbol{x})$ is the corrected response $\boldsymbol{R}_c(f_0, \boldsymbol{\theta})$ obtained by utilizing the DPSS tapers in accordance to the vector of setup parameters $\boldsymbol{x} = [x_1\ x_2]^T = [n\ s]^T$. $\boldsymbol{R}_f = \boldsymbol{R}_f(f_0, \boldsymbol{\theta})$ is the reference performance figure (here, the radiation pattern) obtained from simulations of the AUT EM model. The factor $\alpha = (\boldsymbol{R}_f^T\boldsymbol{R}_f)^{-1}\boldsymbol{R}_f^T\boldsymbol{R}_c(\boldsymbol{x})$ offers an analytical solution for a curve fitting problem that minimizes the difference between $\boldsymbol{R}_c(\boldsymbol{x})$ and $\boldsymbol{R}_f$ through appropriate scaling of the latter. The DPSS-based functional landscape resulting from (3) is highly multimodal which hinders its reliable optimization. Here, an exhaustive search oriented towards evaluation the objective function (3) for a set of designs $\boldsymbol{X} = \{\boldsymbol{x}_p\}_{1 \leq p \leq P}$, followed by domination-based ranking of the obtained responses (so as to identify $\boldsymbol{X}_{opt} \subset \boldsymbol{X}$ that correspond to local minima) is performed. The final result $\boldsymbol{R}_c^*$ is obtained as an average of solutions $\boldsymbol{R}_c(\boldsymbol{x}_\beta)$, $\beta = 1, 2, \ldots,$ and $\boldsymbol{x}_\beta \in \boldsymbol{X}_{opt}$, representing the $\sigma$-quantile of all locally optimal responses, where $\sigma = 0.1$.

## 3    Correction Results

The considered correction framework has been validated based on a total of four experiments conducted in a non-anechoic test site, i.e. an office room ($5.5 \times 4.5 \times 3.1$ m$^3$) that is considered not appropriate for far-field experiments (cf. Fig. 1(a)). The AUT used in the experiments is a spline-based monopole depicted in Fig. 1(b), whereas the RA for the two-antenna system is a Vivaldi structure of Fig. 1(c). The test setup is configured with an angular resolution of 5°. The number of frequency points around $f_0$ is set to $K = 201$ with bandwidth of $B = 3$ GHz. The AUT responses have been measured in the yz-plane at the specified frequencies of interest $f_0 \in \{4, 5, 6, 7\}$ GHz. For all of the considered experiments, the correction performance between the non-anechoic and AC-based measurements is quantified using the root-mean-square error, $e_R$ expressed in decibels [9].

The multitaper post-processing is independently applied at each $f_0$ and optimized according to the exhaustive search procedure outlined in Section 2.3. The correction process via DPSS taper functions is executed based on the set of test designs $\boldsymbol{X}$ with the lower- and upper-bounds of $\boldsymbol{l}_b = [101\ 0.1x_1]^T$ and $\boldsymbol{u}_b = [901\ 0.9x_1]^T$, respectively. The $\boldsymbol{R}_c(\boldsymbol{X})$ responses are ranked based on their corresponding $\boldsymbol{U}(\boldsymbol{X})$ values and the final solution is determined as the average of $\boldsymbol{R}_c(\boldsymbol{x}_\beta)$ responses with the highest rank among the corrected measurements. The process involves the use of five designs, corresponding to the $\sigma$-quantile of the best locally optimal solutions found as a result of exhaustive search of Section 2.3 for $\sigma = 0.1$, and is repeated at all of the considered $f_0$ frequencies.

Figure 2 compares the radiation pattern characteristics of the antenna at the selected frequencies before and after multitaper-based post-processing. The obtained responses demonstrate a significant improvement of non-anechoic measurements fidelity due to correction. Additionally, the antenna performance characteristics before and after refinement, for all considered $f_0$ frequencies, are summarized in Table 1. The average improvement of the radiation pattern fidelity—expressed in terms of $\Delta = |e_R(\boldsymbol{R}_c^*(f_0, \boldsymbol{\theta})) - e_R(\boldsymbol{R}_u(f_0, \boldsymbol{\theta}))|$—is nearly 13 dB. The maximum and minimum $\Delta$ changes are 15.4 dB at 4 GHz and 11 dB at 5 GHz, respectively.

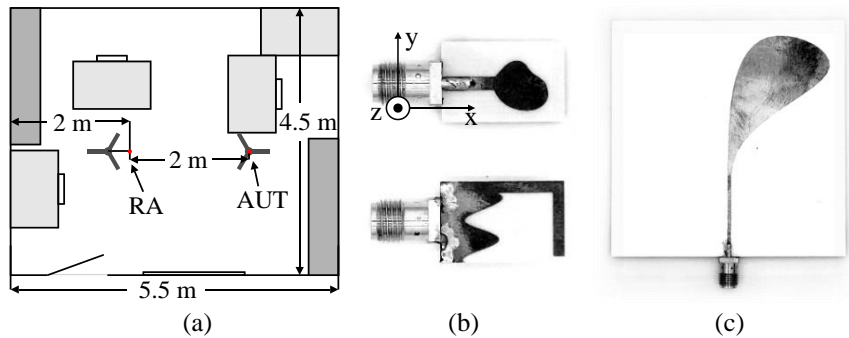

(a)          (b)          (c)

**Fig. 1.** A non-anechoic test site considered for experiments: (a) schematic view highlighting the location of rotary towers with tall (dark gray) and short (light gray) furniture, as well as photographs of (b) the spline-based monopole used as the AUT, and (c) the Vivaldi radiator used as the RA. Note that the considered antenna structures are not in scale.

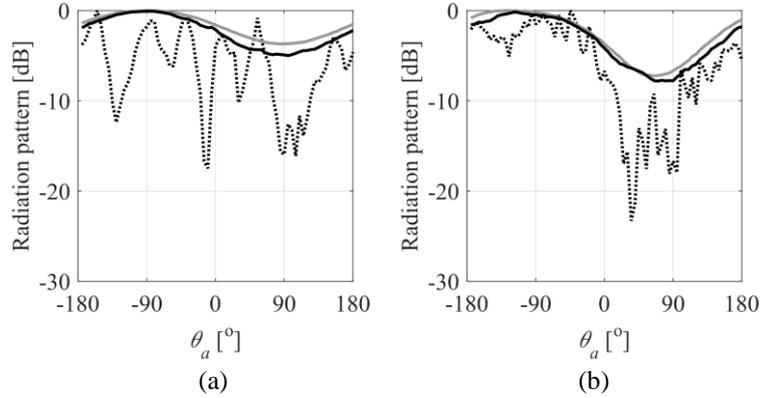

**Fig. 2.** Radiation patterns in yz-plane (cf. Fig. 1) obtained from AC (gray) and the non-anechoic measurements (black) before (⋯), and after correction (—) at: (a) 4 GHz and (b) 7 GHz frequencies.

## 4      Discussion and Comparisons

The discussed approach was compared against other state-of-the-art post-processing algorithms based on time-domain analysis. The selected benchmark methods—implemented for the same configuration parameters ($f_0$, $B$, $K$, $N$)—utilize: (i) manual estimation of RA-AUT distance (i.e., based on physical measurements of the distance) with a rectangular window function, (ii) cognition-based analysis of the impulse response with a Hann function, and (iii) modification of the interval identified based on relative thresholds using a composite window [8], [9], [15]. The correction performance, assessed by the averaged $e_R$ factor, for all considered methods is presented in Table 2. The considered multitaper approach consistently outperforms other methods (for the considered antenna and test frequencies), with improvements in the refined radiation characteristics ranging from 7.2 dB to 10.3 dB, respectively. Additionally, unlike approaches (i) and (ii), the considered method does not rely on expert knowledge nor is dependent on predefined thresholds, as is the case in (iii). The exhaustive search of the setup parameters enhances flexibility in terms of tuning the multitaper post-processing, potentially extending support for a wider range of antenna structures and test sites.

**Table 1.** Correction performance of the multitaper method in non-anechoic test site.

| $f_0$ [GHz] | 4 | 5 | 6 | 7 |
|---|---|---|---|---|
| $e_R(\boldsymbol{R}_u)$ [dB] | −8.8 | −14.1 | −9.9 | −12.8 |
| $e_R(\boldsymbol{R}_c^*)$ [dB] | −24.2 | −25.1 | −22.5 | −25.6 |
| $\Delta$ [dB] | 15.4 | 11.0 | 12.6 | 12.8 |

$\Delta = |e_R(\boldsymbol{R}_c^*(f_0, \boldsymbol{\theta})) - e_R(\boldsymbol{R}_u(f_0, \boldsymbol{\theta}))|$

**Table 2.** Correction performance of the chosen methods in non-anechoic test site.

| Root-mean-square error $e_R$, averaged over the frequencies of interest | | | |
|---|---|---|---|
| (i) | (ii) | (iii) | This work |
| −14.0 dB | −15.9 dB | −17.1 dB | −24.3 dB |

## 5    Conclusion

In this paper, we consider a multitaper post-processing framework for correcting antenna measurements performed in non-anechoic conditions. The method improves RA-AUT transmission obtained from one-shot experiments using mutually orthogonal DPSS taper functions that enable the extraction of the desirable part of the signal from interferences and noise. Demonstrated based on four measurements of a spline-based monopole structure in a non-anechoic test site, the method enhances the fidelity of the non-anechoic responses by an average of nearly 13 dB contrasted to uncorrected characteristics. It has been favorably compared to other techniques applicable to time-domain post-processing of far-field responses.

Future work will focus on enhancing the approach to enable automatic calibration of all relevant setup parameters and development of the ensemble method dedicated to further augment the information embedded within the one-shot measurements.

**Acknowledgments.** This work was supported in part by the National Science Centre of Poland Grant 2021/43/B/ST7/01856, National Centre for Research and Development Grant NOR/POLNOR/HAPADS/0049/2019-00, and Gdansk University of Technology (Excellence Initiative - Research University) Grant 16/2023/IDUB/IV.2/EUROPIUM.

**Disclosure of Interests.** The authors have no competing interests to declare that are relevant to the content of this article.